\documentclass{article}

\usepackage{siunitx}

\usepackage[utf8]{inputenc} 
\usepackage[T1]{fontenc}    
\usepackage{hyperref}       
\usepackage{url}            
\usepackage{booktabs}       
\usepackage{amsfonts}       
\usepackage{nicefrac}       
\usepackage{microtype}      
\usepackage{lipsum}		
\usepackage{graphicx}
\usepackage{doi}

\RequirePackage[utf8]{inputenc}
\RequirePackage[english]{babel}
\RequirePackage{amsmath,amsfonts,amssymb}
\RequirePackage{lmodern}
\RequirePackage[scaled]{helvet}
\RequirePackage[T1]{fontenc}
\RequirePackage{lettrine}

\RequirePackage{authblk}
\RequirePackage{graphicx,xcolor}
\RequirePackage{colortbl}
\RequirePackage{booktabs}
\RequirePackage{algorithm}
\RequirePackage[noend]{algpseudocode}
\RequirePackage{changepage}
\RequirePackage[twoside,%
				letterpaper,includeheadfoot,%
				layoutsize={8.125in,10.875in},%
                layouthoffset=0.1875in,%
                layoutvoffset=0.0625in,%
                left=38.5pt,%
                right=43pt,%
                top=43pt,
                bottom=32pt,%
                headheight=0pt,
                headsep=10pt,%
                footskip=25pt,
                marginparwidth=38pt]{geometry}
\RequirePackage[labelfont={bf,sf},%
                labelsep=period,%
                figurename=Fig.]{caption}

\RequirePackage[numbers,sort&compress,merge]{natbib}
\RequirePackage{jabbrv}
 
\RequirePackage{fancyhdr}  
\RequirePackage{lastpage}  

\newcommand{\mi}{\mathrm{i}}
\newcommand{\me}{\mathrm{e}}
\DeclareSIUnit\bar{bar}

\title{Efficient generation of high-order harmonics in gases}

\date{}

\author[a,*]{R. Weissenbilder}
\author[a]{S. Carlstr\"om}
\author[b]{L. Rego}
\author[a]{C. Guo}
\author[c,d,e]{C. M. Heyl}
\author[f]{P. Smorenburg}
\author[g]{E. Constant}
\author[a]{C. L. Arnold}
\author[a]{A. L'Huillier}

\affil[a]{Department of Physics, Lund University, SE-221 00 Lund, Sweden}
\affil[b]{Grupo de Investigación en Aplicaciones del Láser y Fotónica, Departamento de Física Aplicada, University of Salamanca, Salamanca, Spain}
\affil[c]{Deutsches Elektronen-Synchrotron DESY, Notkestraße 85, 22607 Hamburg, Germany}
\affil[d]{Helmholtz-Institute Jena, Fröbelstieg 3, 07743 Jena, Germany}
\affil[e]{GSI Helmholtzzentrum für Schwerionenforschung GmbH, Planckstraße 1, 64291 Darmstadt, Germany}
\affil[f]{ASML Research, ASML Netherlands B.V., 5504 DR Veldhoven, The Netherlands}
\affil[g]{Université Claude Bernard Lyon 1, CNRS, Institut Lumière Matière, F-69622, Villeurbanne, France}
\affil[*]{Corresponding author: robin.weissenbilder@fysik.lth.se}

\begin{document}

\maketitle

\begin{abstract}
High-order harmonic generation (HHG) in gases leads to short-pulse extreme ultraviolet (XUV) radiation useful in a number of applications, for example, attosecond science and nanoscale imaging. However, this process depends on many parameters and there is still no consensus on how to choose the target geometry to optimize the source efficiency. Here, we review the physics of HHG with emphasis on the macroscopic aspects of the nonlinear interaction. We analyze the influence of medium length, pressure, position of the medium and intensity of the driving laser on the HHG conversion efficiency (CE), using both numerical modelling and analytical expressions. We find that efficient high-order harmonic generation can be realized over a large range of pressures and medium lengths, if these follow a certain hyperbolic equation. The spatial and temporal properties of the generated radiation are, however, strongly dependent on the choice of pressure and medium length. Our results explain the large versatility in gas target design for efficient HHG and provide design guidance for future high-flux XUV sources.
\end{abstract}

\section{Introduction}

Extreme Ultraviolet light sources based on high-order harmonic generation in gases \cite{McPhersonJOSAB1987,FerrayJPB1988} are becoming ubiquitous in many areas of science, from atomic and molecular physics to condensed matter physics, as well as more applied topics, such as coherent imaging \cite{ZurchSR2014} and microscopy \cite{SchmidtAPB2002}. The radiation consists of a train of extremely short light bursts, in the attosecond range, allowing for outstanding temporal resolution. HHG has opened the field of attosecond science, capturing ultrafast electron dynamics in matter \cite{KrauszRMP2009}. The spatial properties of HHG sources are also interesting for some applications since the radiation is spatially-coherent over a broad spectral range. 
During many years, amplified femtosecond titanium-sapphire lasers have been the ``standard'' laser for HHG. Recently, there is an increased diversity of HHG sources driven by a variety of lasers ranging from high energy lasers at low repetition rate, with up to hundreds of mJ energy per pulse \cite{CoudertAS2017,KuehnJoPBAMaOP2017,Kleine2019JPCL,TsatrafyllisSR2016,FuCP2020}, to high average power lasers, based upon optical parametric amplification or simply high-power oscillators, with pulse energies in the \si{\micro\joule} range or below \cite{RothardtNJP2014,CombyOE2019,Mikaelsson2020}. Also, high-power, compact, HHG sources based on post-compressed, ytterbium-doped femtosecond lasers \cite{BoulletOL2009,HaedrichNP2014} are becoming increasingly interesting for industrial applications. Finally, the use of lasers in the mid infrared range allows the generation of high energy photons, in the soft X-ray range \cite{SheehyPRL1999,PopmintchevScience2012}. HHG sources can thus be vastly different, with parameters such as peak power or repetition rate varying by several orders of magnitude \cite{HeylJPBAMOP2017}.

 The HHG efficiency depends both on the response of a single atom to a strong laser field \cite{KrausePRL1992,CorkumPRL1993,LewensteinPRA1994} and, as in any coherent nonlinear optical process, on the phase matching between the waves emitted by individual atoms in a macroscopic medium \cite{LHuillierPRL1991,BrabecRMP2000,KazamiasPRL2003,GaardeJPB2008,PopmintchevNP2010,ConstantPRL1999}. The single atom response depends on the driving laser intensity, wavelength \cite{PopmintchevScience2015}, and the atomic species, and can, to some extent, be boosted by multicolor schemes, at the cost of more complex optical setups \cite{KimPRL2005,BrizuelaSR2013}. The macroscopic response, on the other hand, can be optimized by choosing an appropriate focusing geometry, medium length and pressure. Quasi-phase matching setups using complex gas target designs have also been developed \cite{ChristovOE2000,Hareli2020}. The versatility of HHG sources can be understood, partly, by a rather simple scaling principle: A given configuration for HHG can be scaled up (or down), if certain relations between input energy, focusing geometry, atomic density and medium length are conserved \cite{HeylO2016,NefedovaJESRP2017}. Similar conversion efficiencies can thus be achieved using quite different laser systems, with energy per pulse differing by several orders of magnitude \cite{TakahashiOL2002,TakahashiJOSAB2003,RudawskiRSI2013,HädrichLSA2015}. 

A remaining question is: given a certain laser energy, how should the focusing geometry, medium length and atomic density be chosen in order to maximize the HHG efficiency? Although HHG sources have been around for more than 30 years, there is no consensus regarding optimal medium design, and current setups vary from high pressure gas jets, to low pressure cells, semi-infinite cells or capillaries, just to cite the most common ones. 

In this work, we investigate the dependence of the HHG conversion efficiency, spatial and temporal properties on the medium length and pressure, with the help of numerical simulations and analytical derivations. After a tutorial review of the main physics behind HHG, we present simulations based on solving the time-dependent Schr\"odinger equation (TDSE) and a wave propagation equation, including dispersion, absorption and ionization, for the 23\textsuperscript{rd} harmonic in argon and the 69\textsuperscript{th} harmonic in neon, using 810\,nm, 22\,fs full width at half maximum (FWHM) laser pulses. We find that for any given focusing geometry, efficient HHG can be achieved over a wide range of pressures, as long as the medium length is appropriately chosen. The relation between pressure and length, corresponding to a high conversion efficiency, follows a hyperbolic equation, which is derived analytically. However, the choice of pressure (or length) strongly influences the temporal and spatial properties of the emitted radiation. Our results, which can easily be extended to other harmonics, laser wavelengths and atomic species, explain why the achieved conversion efficiencies are often comparable despite the large variety of design, ranging from gas jets to capillaries, making HHG such a robust technique.

The article is organized as follows: In Section \ref{sec:interaction}, we give an overview of ionization by strong laser fields and photoionization by extreme ultraviolet light in argon and neon. In Section \ref{sec:principles}, we discuss the physics of high-order harmonic generation in gases, and develop a one-dimensional analytical model to describe phase matching. In Section \ref{sec:simulations}, we present the results of our simulations and compare them to the predictions of the model. In Section \ref{sec:examples}, we give two examples of how to optimally choose the geometry and gas target, and we conclude in Section \ref{sec:discussion}.

\section{Ionization of atoms}
\label{sec:interaction}

\subsection{Atoms in strong laser fields}
\label{ssec:strong}
\begin{figure}[ht]
\begin{center}
\includegraphics{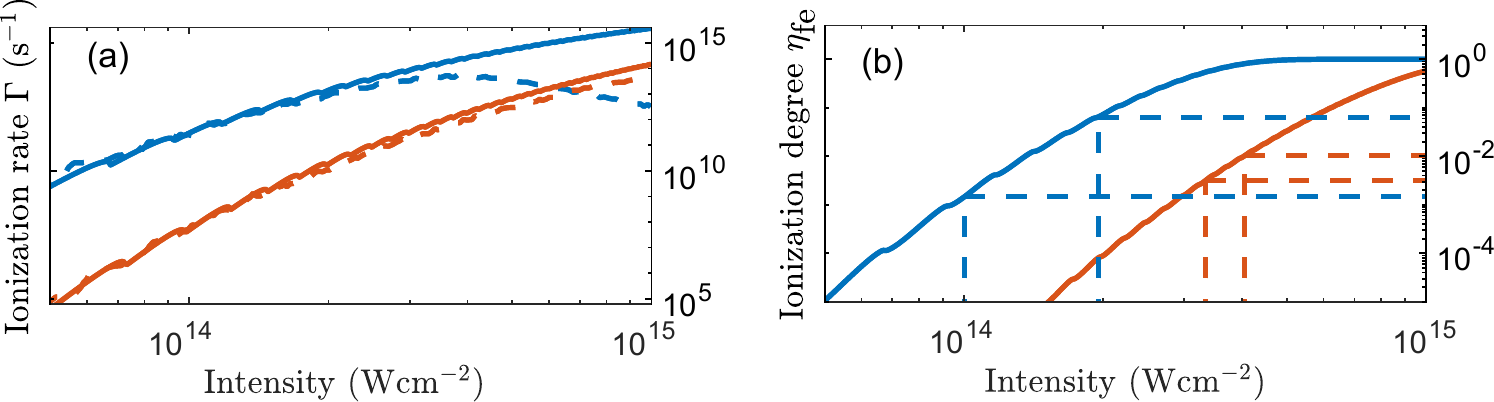}
\caption{(a) Ionization rates, $\Gamma$, as a function of intensity in Ne (red) and Ar (blue), calculated using the Perelomov-Popov-Terent'ev approximation \cite{Perelomov1966} (solid) and solving the time dependent Schrödinger equation \cite{Schafer2009} (dashed). The laser wavelength is 810 nm and (b) Ionization degree as a function of intensity in Ne (red) and Ar (blue), calculated using PPT ionization rates for a Gaussian pulse with 22\,fs FWHM pulse duration. The ionization degree is given at the time $t=0$ corresponding to the peak of the laser pulse. The dashed vertical lines define an intensity window, bounded from below by the cut-off intensity obtained from the three-step model, and from above by a critical intensity (see section \ref{sec:principles}\ref{sec:IntensityWindow}). The dashed horizontal lines show the ionization degree associated with these intensities.}
\label{fig:ionization}
\end{center}
\end{figure}

HHG is intimately related to ionization in strong laser fields, which initiates the single atom response, and at the same time limits the macroscopic yield when a too high density of free electrons in the medium prohibits phase matching and modifies the laser propagation. Fig.~\ref{fig:ionization}(a) shows ionization rates, $\Gamma$, as a function of laser intensity, $I$, for Ar and Ne for a wavelength of 810\,nm. Results obtained by solving the TDSE numerically in the single active electron approximation \cite{Schafer2009} (dashed line) compare well with calculations using the so-called Perelomov-Popov-Terent'ev (PPT) \cite{Perelomov1966} formula (solid line). The deviation at high intensity in the Ar case corresponds to depletion, not included in PPT. The steps shown using the PPT approximation are due to channel closings related to the increase of the ionization energy in strong laser fields. The well known Ammosov-Delone-Krainov (ADK) \cite{AmmosovSPJ1986a} approximation, which is the low frequency limit of PPT, valid in the tunneling limit, can also be used to calculate ionization rates, though the results (not shown) deviate more than PPT from those obtained by solving the TDSE. For few cycle pulses, where sub-cycle dynamics can be important \cite{GeisslerPRA2000}, the Yudin-Ivanov (YI) ionization rate \cite{YudinPRA2001B}, which is an extension to the cycle averaged PPT, can be used.

In Fig.~\ref{fig:ionization}(b) we present the ionization degree ($\eta_{\textrm{fe}}$) obtained after half of the laser pulse, as function of laser intensity, for a Gaussian pulse with 22\,fs pulse duration FWHM. The ionization degree at the peak of the pulse, assumed centered at $t=0$, is given by
\begin{equation}
    \eta_{\textrm{fe}}= 1- \exp{\left(-\int_{-\infty}^{0} \Gamma[I(t)] dt\right)}.
\end{equation}

\subsection{Interaction of XUV radiation with atoms}
\label{ssec:absorption}
\begin{figure}[ht]
  \centering
    \includegraphics{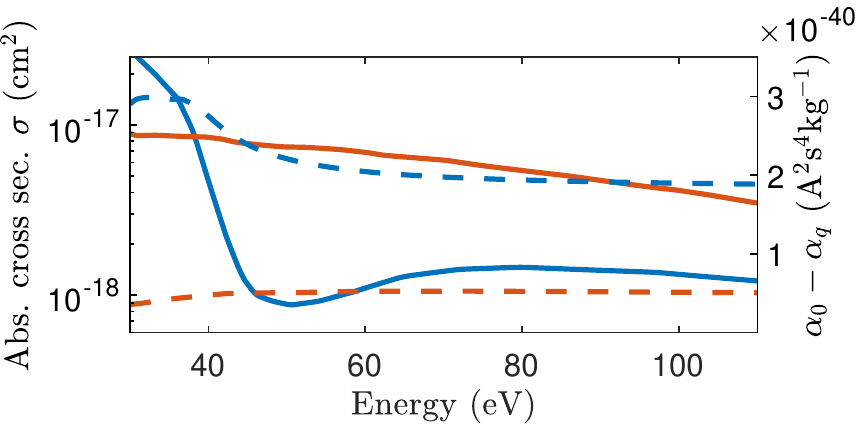}
      \caption{Absorption cross section as a function of photon energy (solid line) in Ar (blue) and Ne (red). Difference between the static polarizability $\alpha_0$ and the polarizability $\alpha_q$ (see \ref{sec:principles}\ref{ssec:phasematching}) as a function of photon energy (dashed line) in Ar (blue) and Ne (red). Data from \cite{HenkeADNDT1993}.}
      \label{fig:absorption}
\end{figure}

The second process intrinsically linked to HHG is photoionization by XUV radiation. Indeed, the reverse phenomenon, recombination, is an inherent step of the single atom response. In addition, the interaction of XUV light with the medium affects propagation, through dispersion and absorption (real and imaginary parts of the refractive index). Fig.~\ref{fig:absorption} shows the absorption cross section, $\sigma$ (solid lines), for Ar and Ne in the energy range reached with HHG using a near infrared laser driver. The rapid variation of the cross section in Ar is due to a Cooper minimum at 52\,eV \cite{CooperPR1962}. The influence of dispersion will be discussed in section \ref{sec:principles}\ref{ssec:phasematching}.

\section{Basic principles of high-order harmonic generation}
\label{sec:principles}

\subsection{Single atom response}
\label{sec:SAR}
The physics of HHG is well described using a semi-classical three-step model \cite{KulanderAILF1992,CorkumPRL1993}, or equivalently a quantum mechanical description based on the Strong-Field-Approximation \cite{LewensteinPRA1994}. Briefly, the first step of the process is ionization by a strong low-frequency field, as described in Section \ref{sec:interaction}\ref{ssec:strong}. The second step is the classical motion of the electron in the continuum driven by the laser field. Most of the electron trajectories do not come back to the nucleus, hence contribute to the partial ionization of the medium. For the trajectories coming back, there is a small probability for recombination back to the ground state. This third step, which is the inverse of the photoionization process described in Section \ref{sec:interaction}\ref{ssec:absorption}, is accompanied by the emission of extremely short, attosecond, XUV light pulses. Interference between attosecond pulses emitted twice per laser cycle, with a sign flip between consecutive pulses, leads to the emission of odd-order harmonics of the laser light. There are mainly two electron trajectories leading to a given XUV photon energy, originating at different tunneling times. These trajectories are referred to as "short" or "long", depending on the electron excursion time.

The three step model successfully predicts the cut-off energy for high-order harmonic generation, equal to
\begin{equation}
    E_\textrm{cut-off} = I_p + 3.17U_p.
    \label{eq:cutoff}
\end{equation}
$I_p$ is the ionization energy and $U_p = e^2|E_0|^2/(4m_e\omega^2)$ is the ponderomotive energy, where $e, m_e$ are the electron charge and mass, $E_0$ is the laser field strength and $\omega$ is the laser frequency.

\subsection{Propagation of high-order harmonics and geometrical scaling}
\label{sec:scaling}
Using the slowly-varying envelope and paraxial approximations \cite{LHuillier1992}, the propagation of the $q^{\textrm{th}}$ harmonic field can be described by a wave equation,
\begin{equation}
\nabla_\perp^2 E_q -2 \mi k_q \frac{\partial E_q}{\partial z} = -\mu_0 q^2\omega^2 P_q \me^{-\mi (qk_1-k_q)z},
\label{eq:prop}
\end{equation}
where the harmonic field and polarization are related to the slowly varying envelopes as $E_q \exp[\mi(q\omega t - k_q z)] + \textrm{c.c.}$ and $P_q \exp[\mi(q\omega t - qk_1z)] + \textrm{c.c.}$ 
The symbol $\nabla^2_\perp$ refers to double differentiation with respect to the transverse directions $x$ and $y$ ($z$ denoting the propagation direction) and $\mu_0$ is the vacuum permeability. The wavevector of the fundamental field is $k_1= n_1 \omega/c$, and the wavevector of the $q^{\textrm{th}}$ harmonic field is $k_q=n_q q\omega/c$, where $c$ is the speed of light and $n_1$ and $n_q$ are the refractive indices at frequencies $\omega$ and $q\omega$ respectively. For propagation distances small compared to the Rayleigh length of the driving laser, diffraction, {\it i.e.} the first term of \eqref{eq:prop}, may be neglected, resulting in a one-dimensional wave equation. Integrating the one-dimensional equation for a homogeneous medium with length $L$ and excluding the effect of reabsorption yields
\begin{equation}
|E_q|^2 \propto \left|\frac{\sin(\Delta k L/2)}{\Delta k}\right|^2, 
\label{eq:noabs}
\end{equation}
where $\Delta k= qk_1-k_q$ is the phase mismatch between the polarization field at frequency $q\omega$, induced by the response of the medium to the fundamental field, and the generated harmonic field. This highlights the importance of minimizing $\Delta k$ to optimize the conversion efficiency. \eqref{eq:noabs} can be generalized to include diffraction and focusing, assuming a power law (not necessarily within lowest-order perturbation theory) for the single atom response \cite{LHuillierJPB1991}.

Under the transformation ($x$, $y$, $z$, $\rho$) $\rightarrow$ ($\eta x$, $\eta y$, $\eta^2 z$, $\rho/\eta^2$), where $\eta$ is a scaling factor \cite{HeylO2016}, we have
\begin{equation}
    \nabla_\perp^2  \rightarrow \frac{\nabla_\perp^2}{\eta^2},\ \   \ 
    \frac{\partial}{\partial z}\rightarrow \frac{\partial}{\eta^2\partial z}, \ \ \ 
    P_q \rightarrow \frac{P_q}{\eta^2},
\end{equation}
so that Eq.~(\ref{eq:prop}) remains invariant. Scaling $z$ implies to scale the medium length by the same factor. Such a scaling also requires to keep the same laser intensity distribution in the medium, and therefore to scale the input (and output) energies by $\eta^2$. This geometrical scaling allows us to discuss the efficiency of HHG independently of the focusing geometry. The results presented below will thus be scaled by the Rayleigh length $z_R$.

\subsection{Phase mismatch in HHG}
\label{ssec:phasematching}
In HHG in gases, there are four contributions to the phase mismatch between the fundamental field and the generated q$^\textrm{th}$ harmonic field. These contributions are due to the dispersion in the neutral medium $\Delta k_{\text{at}}$, the presence of free electrons $\Delta k_{\textrm{fe}}$, the influence of the laser focusing $\Delta k_{\text{foc}}$ and finally the so-called dipole phase contribution $\Delta k_{\text{i}}$. The dipole phase is due to the single atom response \cite{LewensteinPRA1995}, and depends on the electron trajectory in the continuum (short or long). In this section, we discuss the phase mismatch along the propagation axis due to these four contributions, neglecting the contribution of ions \cite{PopmintchevScience2015},
\begin{equation}
    \Delta k =  \Delta k_{\text{at}} + \Delta k_{\textrm{fe}} + \Delta k_{\text{foc}} + \Delta k_{\text{i}}.
\end{equation}
The dispersion in the neutral medium is equal to 
\begin{equation}
    \Delta k_{\text{at}}= (n_1-n_q)\frac{q\omega}{c}.
    \label{eq:neutral}
\end{equation} 
At the fundamental frequency, $n_1>1$ while at the harmonic frequencies above the ionization threshold $n_q<1$, so that this contribution is positive. We introduce the polarizability $\alpha_q=2\epsilon_0 (n_q-1)/\rho$ at frequency $q\omega$, where $\epsilon_0$ is the vacuum permittivity and $\rho$ is the atomic density of the medium. Eq.~(\ref{eq:neutral}) can be written as a function of the polarizabilities $(\alpha_1)$ and $(\alpha_q)$ as
\begin{equation}
    \Delta k_{\text{at}}= \frac{q\omega \rho}{2\epsilon_0c}\left(1-\eta_\textrm{fe}\right)\left(\alpha_1-\alpha_q\right),
    \label{eq:neutral3}
\end{equation} 
where $\eta_\textrm{fe}$ is the ionization degree in the medium. The polarizability at the fundamental wavelength ($\alpha_1$) can be approximated by the static polarizability ($\alpha_0$), and the ionization degree can be assumed to be small ($\eta_\textrm{fe} \ll 1$), so that
\begin{equation}
    \Delta k_{\text{at}}\approx \frac{q\omega \rho}{2\epsilon_0c}\left(\alpha_0-\alpha_q\right) > 0.
    \label{eq:neutral2}
\end{equation} 
The variation of $\alpha_0-\alpha_q$ as a function of photon energy is plotted in Fig.~\ref{fig:absorption} (dashed lines) in Ar and Ne. In the energy region shown in Fig.~\ref{fig:absorption}, the refractive index in Ar and Ne is below 1 and $\alpha_0-\alpha_q \geq 0$.

For a guided geometry, {\it e.g.} a hollow core capillary, the phase mismatch should also include the effect of the mode dispersion in the wave guide \cite{RundquistScience1998}, leading to an additional phase mismatch
\begin{equation}
   \Delta k_{\textrm{wg}} = - q\frac{u_{nm}^2\lambda}{4\pi a^2},
\end{equation}
where $a$ is the radius of the wave guide, $\lambda$ is the laser wavelength, and $u_{nm}$ are constant factors of the propagating modes, given by the $m^\textrm{th}$ root of the equation $J_{n-1}(u_{nm})=0$ \cite{MarcatiliBSTJ1964}, where $J_n(x)$ is the Bessel function of the first kind of order $n$. Typically HHG in guided geometries use the first order mode, for which $u_{11}=2.4$. 

The dispersion due to free electrons can be expressed as
\begin{equation}
\Delta k_{\textrm{fe}}= -\frac{q\omega\rho}{2\epsilon_0c}\frac{\eta_\textrm{fe}e^2}{m_e} \left(\frac{1}{\omega^2} - \frac{1}{q^2\omega^2}\right) < 0.
\label{eq:fe}
\end{equation}
The second term in the parenthesis above, which scales as $1/q^2$, can be neglected for high-order harmonics.

When a laser beam goes through a focus, the phase varies (in addition to the usual $kz$). For a Gaussian beam the phase variation is the Gouy phase shift $\zeta(z)=-\tan^{-1}(z/z_R)$, where $z_R$ is the Rayleigh length. The Gouy phase leads to an increase of the fundamental phase velocity. Neglecting the effect of the focusing of the harmonic beam, the phase mismatch is
\begin{equation}
\Delta k_\text{foc} = q\frac{d\zeta}{dz} = -\frac{qz_R}{z^2+z_R^2} < 0.
\label{eq:gouy}
\end{equation}
$\Delta k_\text{foc}$ is negative and equal to $-q/z_R$ when $z=0$. For a guided geometry $\Delta k_\text{foc}$ is equal to zero.

\begin{figure}[ht]
\begin{center}
\includegraphics{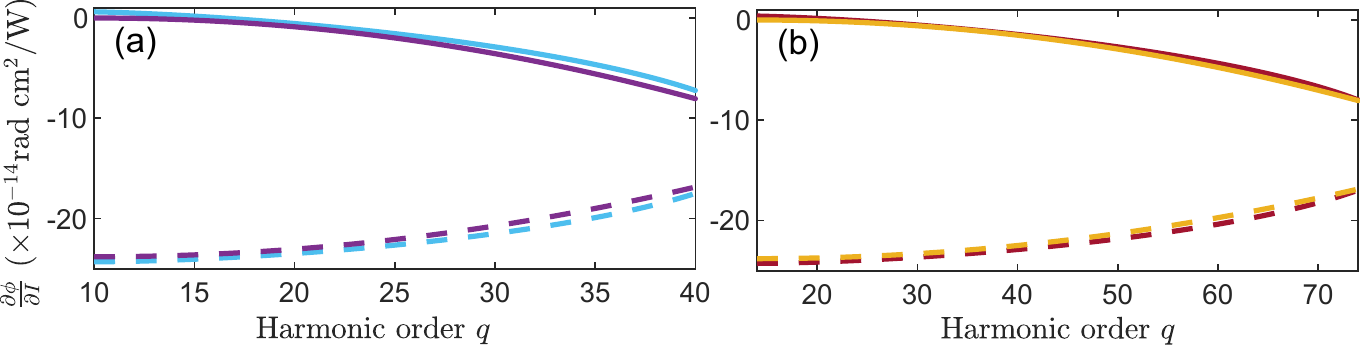}
\caption{$\partial\Phi_\textrm{i}/\partial I$ [see \eqref{eq:alphai}] in (a) Ar, with $I = 2.5\times10^{14}$\,Wcm$^{-2}$ and (b) Ne with $I = 5\times10^{14}$\,Wcm$^{-2}$ for the short (solid) and long (dashed) trajectories, obtained by solving the saddle point equations in the strong field approximation [blue, red] and from the model in \eqref{eq:phasef} [purple, yellow].}
\label{fig:alpha}
\end{center}
\end{figure}

Finally, to estimate the phase mismatch $\Delta k_{\text{i}}$, we use an approximate expression for the phase accumulated by the electron in the continuum \cite{GuoJPBAMOP2018},
\begin{equation} 
\Phi_\mathrm{i}=\alpha_\mathrm{i}I+t_\mathrm{pi}\left(q\omega-\tfrac{I_p}{\hbar}\right)+ \frac{\gamma_\mathrm{i}}{I}\left(q\omega-\tfrac{I_p}{\hbar}\right)^2,
\label{eq:phasef}
\end{equation}
where i=s,$\ell$ for the short and long trajectories respectively, and $\hbar$ is the reduced Planck constant. In the literature, the dipole phase is commonly expressed as $\Phi_\textrm{i}=\alpha_\textrm{i}(q\omega)I$ \cite{LewensteinPRA1995}. The advantage of \eqref{eq:phasef}, where $\alpha_i I$ now refers only to the first harmonic above the ionization threshold, is that it gives a simple analytical expression for the frequency dependence of the phase. For the short trajectory $\alpha_\mathrm{s}=0$ and for the long trajectory $\alpha_\mathrm{\ell} = -0.16\alpha\lambda^3/(m_\mathrm{e}c^3)$ 
\cite{WikmarkPNAS2019}, with $\alpha$ being the fine structure constant. The second term, where $t_\mathrm{pi}$ is an approximate threshold return time, represents the group delay and does not influence phase matching. The third term describes the group delay dispersion of the attosecond pulse. The quantity $\gamma_\mathrm{s}$ is equal to $0.22cm_\mathrm{e}/(\alpha\lambda)$, while $\gamma_\mathrm{\ell}=-0.19cm_\mathrm{e}/(\alpha\lambda)$. 
Fig.~\ref{fig:alpha} compares the predictions of this model for
\begin{equation}
\frac{\partial\Phi_\mathrm{i}}{\partial I}=  \alpha_\mathrm{i}-\gamma_\mathrm{i}\left(q\omega - \tfrac{I_p}{\hbar} \right)^2\frac{1}{I^2},  
\label{eq:alphai}
\end{equation}
with calculations performed by solving the saddle point equations within the Strong Field Approximation \cite{LewensteinPRA1995,VarjuJMO2005}. The good agreement between the two calculations validates the analytical model which will be used throughout.

The dependence of $\Phi_\mathrm{i}$ with intensity, and therefore with $z$, leads to a phase mismatch
\begin{equation}
\Delta k_{\text{i}} = \frac{\partial\Phi_\mathrm{i}}{\partial I}\ \frac{\partial I}{\partial z} 
\label{eq:dip}
\end{equation}
 $\Delta k_{\text{i}}$ changes sign across the focus, being negative when $z \le 0$ and positive when $z \ge 0$. For a Gaussian beam characterized by a Rayleigh length $z_R$, and a laser focus at $z = 0$, \eqref{eq:dip} becomes
\begin{equation}
\Delta k_{\text{i}} =-\frac{2z\beta_\textrm{i}(z)}{z^2+z_R^2},
\label{eq:dki}
\end{equation}
where
\begin{equation}
    \beta_\textrm{i}(z) = \alpha_\mathrm{i}I(z)- \frac{\gamma_\mathrm{i}}{I(z)}\left(q\omega - \tfrac{I_p}{\hbar}\right)^2 <0.
    \label{eq:betai}
\end{equation}
In a guided geometry the intensity is constant and thus $\Delta k_\textrm{i}=0$.

\subsection{Phase matching on axis}
\label{sec:PhaseMatchPressure}
Phase matching amounts to compensating the effect of the fundamental Gouy phase (and to some extent the $q^\textrm{th}$ harmonic dipole phase) by choosing appropriately the pressure-dependent refractive indices at frequencies $\omega$ and $q\omega$ for a given ionization degree.

\begin{figure}[ht]
  \centering
    \includegraphics[width=0.98\linewidth]{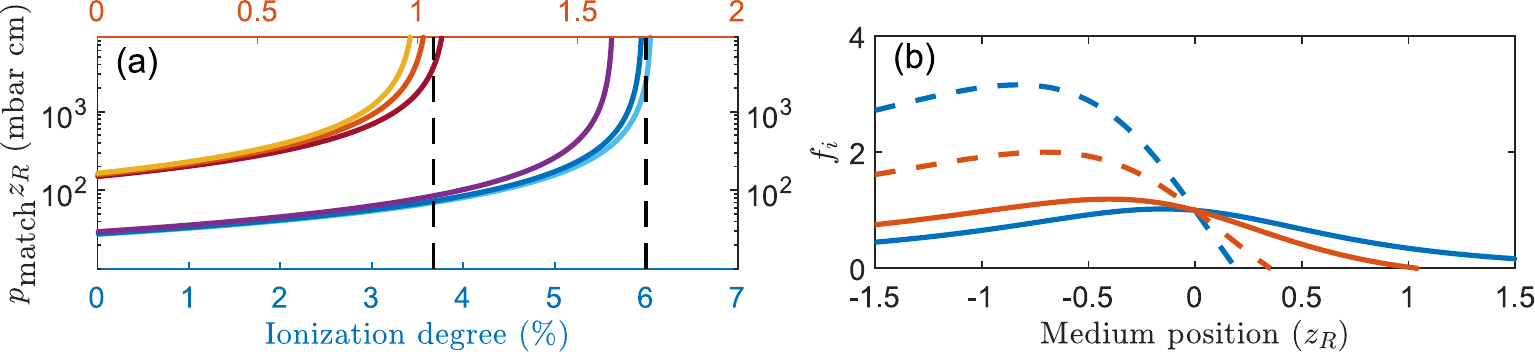}
      \caption{(a) Phase matching pressure as a function of ionization degree in Ar for different harmonic orders [$q$ = 19, 23, 27 (light blue, blue, purple)] and in Ne for different harmonic orders [$q$ = 61, 69, 77 (dark red, red, yellow)]. The dashed black lines correspond to the critical ionization ($\eta_\textrm{fe}^\textrm{mac}$) for the 23$^{\textrm{rd}}$ and 69$^{\textrm{th}}$ harmonic in Ar and Ne respectively. The ionization degree is indicated with the bottom (top) $x$-axis for Ar (Ne). (b) Factor $f_\textrm{i}$ [see \eqref{eq:pmatch4}] describing the effect of moving the medium relative to the laser focus for the 23$^{\textrm{rd}}$ harmonic in Ar and the 69$^{\textrm{th}}$ harmonic in Ne for the long (dashed) and short (solid) trajectory. The intensities are chosen to be 2.5 $\times 10^{14}\,$\si{\watt\per\centi\meter\squared}, and 5 $\times 10^{14}\,$\si{\watt\per\centi\meter\squared}, respectively in the medium.}
      \label{fig:pmatch}
\end{figure}
 
We first consider the case where the nonlinear medium is centered at the focus of the laser, so that $\Delta k_{\textrm{i}}$ can be neglected. Introducing $\Delta \kappa_\text{at}= \Delta k_\text{at}/\rho$ and $\Delta \kappa_\textrm{fe}= \Delta k_\textrm{fe}/(\rho\eta_\textrm{fe})$, the atomic density required to achieve phase matching ({\it i.e.} $\Delta k = 0$) is    
\begin{equation}
    \rho_\text{match} = - \frac{\Delta k_\textrm{foc}}{\Delta \kappa_\text{at}+\eta_\textrm{fe}\Delta \kappa_\textrm{fe}}.
    \label{eq:rhomatch}
\end{equation}
The atomic density is related to the pressure through the ideal gas law $p = \rho k_{\textrm{B}}T$, where $k_{\textrm{B}}$ is the Boltzmann constant and $T$ is the temperature in Kelvin, often assumed to be the room temperature. Although not always true, {\it e.g.} for a gas jet with supersonic expansion, we will here express our results in terms of pressure and not atomic density, assuming that these are related by the ideal gas law. Using $\Delta k_\text{foc}=-q/z_R$, we have,
\begin{equation}
    p_\text{match}z_R = \frac{qk_{\textrm{B}}T}{\Delta \kappa_\text{at}+\eta_\textrm{fe}\Delta \kappa_\textrm{fe}},
    \label{eq:pmatch}
\end{equation}
where $p_\textrm{match}$ is the phase matching pressure. This equation holds when $\Delta\kappa_\textrm{at} > |\eta_{\textrm{fe}}\Delta \kappa_\textrm{fe}|$, requiring that the ionization degree is less than a critical value $\eta_{\textrm{fe}}^\textrm{mac}$ defined as \cite{RundquistScience1998,RundquistPRL1999}, 
    \begin{equation}
        \eta_\textrm{fe}^\textrm{mac} = \frac{m_\textrm{e}\omega^2}{e^2}\left(\alpha_0 -\alpha_q \right),
        \label{eq:etafemac}
    \end{equation}
    where $\alpha_0 -\alpha_q$ is obtained from Fig.~\ref{fig:absorption}. The critical ionization degree, $\eta_\textrm{fe}^\textrm{mac}$, corresponds to equal phase velocities of the fundamental and $q^\text{th}$ harmonic fields. An ionization degree above $\eta_\textrm{fe}^\textrm{mac}$ implies that the phase velocity difference changes sign and that phase matching on axis is no longer possible. Eq.~(\ref{eq:pmatch}) can be rewritten as a function of the ionization degree as
    \begin{equation}
         p_\text{match}z_R = \frac{2m_\textrm{e}\omega\epsilon_0c k_{\textrm{B}}T}{e^2(\eta_\textrm{fe}^\textrm{mac}-\eta_\textrm{fe})}.
    \label{eq:pmatch3}
    \end{equation}

Fig.~\ref{fig:pmatch}(a) shows $p_\textrm{match}z_R$ in \si{\milli\bar\centi\meter} (or equivalently in \si{\pascal\meter}) as a function of the ionization degree in the medium for different harmonic orders in Ar and in Ne. The critical ionization degree, indicated by the dashed lines, is about 6$\%$ for the 23$^{\textrm{rd}}$ harmonic in Ar, while for the 69$^{\textrm{th}}$ harmonic in Ne it is slightly more than 1.1$\%$. 

When the medium is not centered at the laser focus, we take into account $\Delta k_\textrm{i}$ and generalize Eq.~(\ref{eq:pmatch3}) to
\begin{equation}
         p_\text{match}z_R = \frac{2m_\textrm{e}\omega\epsilon_0c k_{\textrm{B}}Tf_\textrm{i}}{e^2(\eta_\textrm{fe}^\textrm{mac}-\eta_\textrm{fe})},
    \label{eq:pmatch5}
    \end{equation}
    where
\begin{equation}
\begin{aligned}
 f_\textrm{i} \!& =\!\frac{z_R^2}{z^2+z_R^2}\left(1+\frac{2z\beta_\textrm{i}(z)}{qz_R}\right).
\end{aligned}
\label{eq:pmatch4}
\end{equation}
Here, $z$ is the position of the center of the medium relative to the laser focus ($f_\textrm{i} = 1$ for $z = 0)$. This factor is shown in Fig.~\ref{fig:pmatch}(b) for the 23$^{\textrm{rd}}$ and 69$^{\textrm{th}}$ harmonic in Ar and Ne for intensities $I = 2.5 \times 10^{14}\,$\si{\watt\per\centi\meter\squared}, and $I = 5 \times 10^{14}\,$\si{\watt\per\centi\meter\squared}, respectively, in the center of the medium.

While $f_\textrm{i}$ remains close to unity for the short trajectory, it displays a strong variation for the long trajectory. Consequently, phase matching will be achieved at different pressures for the two trajectories when the medium is moved relative to the laser focus. In addition, $f_\textrm{i}$ becomes negative for the long trajectory when the medium is located after the laser focus, preventing it from being phase matched in conditions where $\eta_\textrm{fe}<\eta_\textrm{fe}^\textrm{mac}$ \cite{AntoinePRL1996}. However, in this case, phase matching above the critical ionization might become possible.

\subsection{Intensity window for efficient HHG}
\label{sec:IntensityWindow}
Efficient HHG is limited by the intensity and duration of the driving laser pulse, through the cut-off energy [see \eqref{eq:cutoff}], above which the harmonic yield decays rapidly, and through the critical ionization degree [see \eqref{eq:etafemac} and Fig.~\ref{fig:pmatch}(a)], above which phase matching cannot be achieved. 
This allows us to define an intensity window for efficient HHG. The minimum intensity $I_\textrm{mic}$ is the cut-off intensity, obtained from \eqref{eq:cutoff}, 
\begin{equation}
    I_\textrm{mic} = \frac{m_e\omega^2}{3.17\times2\pi\alpha}\left(q\omega-\frac{I_p}{\hbar}\right),
\end{equation}
where the laser intensity $I$ is related to the field strength $E_0$ by $I= c\epsilon_0|E_0|^2/2$. The index ``mic'' indicates that this intensity cut-off is related to the microscopic single atom response. Using Fig.~\ref{fig:ionization}(b), we can deduce the corresponding ionization degree $\eta_\textrm{fe}^\textrm{mic}$. 

\begin{figure}[ht]
  \centering
    \includegraphics[scale=0.9]{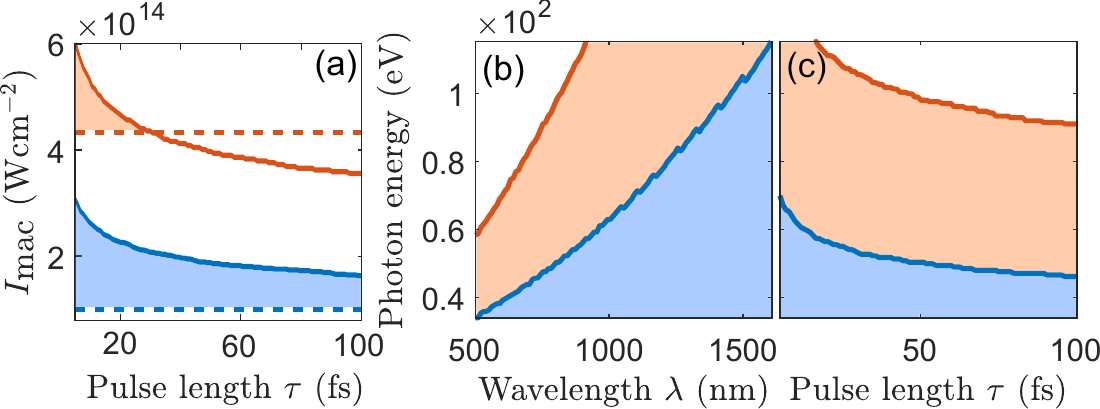}
      \caption{(a) Variation of $I_\textrm{mac}$ as a function of FWHM pulse duration for $\lambda=810$\,nm for the 23$^\textrm{rd}$ harmonic in Ar (blue solid) and 69$^\textrm{th}$ harmonic in Ne (red solid). The horizontal dashed lines indicate the corresponding value of $I_\textrm{mic}$. (b) Variation of the cut-off harmonic (such that $I_\textrm{mic}=I_\textrm{mac}$) as a function of driving wavelength for an 8-cycle FWHM pulse length. (c) Variation of the cut-off harmonic as a function of pulse duration for a driving wavelength of 810\,nm in Ar (blue) and in Ne (red). The shaded regions indicate where on-axis phase matched HHG is possible.}
      \label{fig:ImacTau}
\end{figure}

For a given laser pulse duration, we introduce $I_\textrm{mac}$, as the peak intensity for which $\eta_\textrm{fe}=\eta_\textrm{fe}^\textrm{mac}$ at the peak of the pulse [as shown in Fig.~\ref{fig:ionization}(b)].
We indicate the values of $I_\textrm{mic}$ and $I_\textrm{mac}$ and the corresponding ionization degrees $\eta_\textrm{fe}^\textrm{mic}$ and $\eta_\textrm{fe}^\textrm{mac}$ for the 23$^{\textrm{rd}}$ harmonic in Ar and the 69$^{\textrm{th}}$ harmonic in Ne in 
Table~\ref{tab:IntensityWindow}, for a 22\,fs FWHM Gaussian pulse. 

For a given generating gas and harmonic order, as the pulse duration increases, $I_\textrm{mac}$ generally decreases, while $I_\textrm{mic}$ is constant, thus reducing the intensity window for efficient HHG. This is shown in Fig.~\ref{fig:ImacTau}(a) for the 23$^{\textrm{rd}}$ harmonic in Ar and the 69$^{\textrm{th}}$ harmonic in Ne, where the shaded regions indicate the pulse lengths and peak intensities for which efficient, phase matched HHG is possible. Similarly, as the harmonic order increases, $I_\textrm{mac}$ generally decreases [see Fig.~\ref{fig:pmatch}(a)], while $I_\textrm{mic}$ increases, thus reducing the intensity window for efficient HHG.

The condition $I_\textrm{mic}= I_\textrm{mac}$ defines a cut-off harmonic \cite{PopmintchevOL2008} above which on-axis phase matched HHG is no longer possible. This cut-off harmonic can be moved to higher XUV photon energies by using longer driving wavelengths \cite{PopmintchevOL2008}, as shown in Fig.~\ref{fig:ImacTau}(b), at the cost of a lower CE \cite{ShinerPRL2009}. Alternatively, as shown in Fig.~\ref{fig:ImacTau}(c), shorter driving pulse lengths can be used \cite{KlasPX2021,PopmintchevNP2010}. This highlights the importance of using short driving laser pulses for efficient harmonic generation, and longer wavelengths to reach the soft X-ray regime \cite{SheehyPRL1999,PopmintchevScience2012}. Using short pulses with very high intensities, it is possible to produce very high photon energies, also for near infrared wavelengths, through a process called nonadiabatic self-phase matching \cite{GeisslerPRA2000,SeresPRL2004}. However, recent experimental comparisons of the efficiency favor long driving wavelengths \cite{ChevreuilOE21}.

\begin{table}[htbp]
\centering
\caption{\bf Intensity window for efficient generation of the $23^\textrm{rd}$ harmonic in Ar and $69^\textrm{th}$ harmonic in Ne, the corresponding ionization degrees, absorption cross sections and $\alpha_0-\alpha_q$. The intensity window is calculated for a pulse duration $\tau=22$\,fs [see Fig.~\ref{fig:ionization}(b)].}
\begin{tabular}{ccc}
 & Ar $q=23$ & Ne $q=69$\\
 \hline
$I_\textrm{mic}$ (\si{\watt\per\centi\meter\squared})& $1.0 \times 10^{14}$ & $3.3\times 10^{14}$ \\
$I_\textrm{mac}$ (\si{\watt\per\centi\meter\squared}) & $2.1 \times 10^{14}$ & $4.6 \times 10^{14}$\\
\hline
$\eta_\textrm{fe}^\textrm{mic} \,(\%)$ & 0.15 & 0.16 \\
$\eta_\textrm{fe}^\textrm{mac} \,(\%)$  & 6.0 & 1.0  \\
\hline
$\sigma_\textrm{abs}$ (\si{\centi\meter\squared})  & $1.5\times 10^{-17}$ & $3.8\times 10^{-18}$ \\
$\alpha_0-\alpha_q$ (\si{\ampere\squared\second\tothe{4}\per\kilo\gram})  & $3.0 \times 10^{-40}$ & $5.1\times 10^{-41}$  \\
\hline
\end{tabular}
  \label{tab:IntensityWindow}
\end{table}

\subsection{Phase matching off-axis}
\label{sec:offAxis}
So far we have only considered phase matching along the propagation direction. Phase matching can also be achieved in other directions, not parallel to the laser propagation. Phase matching can in general be formulated as a vectorial momentum conservation equation \cite{BalcouPRA1997},
\begin{equation}
  \boldsymbol{k}_q=q\boldsymbol{k_1}+ \boldsymbol{K_\text{i}}.   
\end{equation}
Here, $\boldsymbol{k_1}$ and $\boldsymbol{k}_q$ represent the wavevectors of the fundamental and $q^\textrm{th}$ harmonic fields, which include dispersion and focusing, and $\boldsymbol{K_\text{i}}=\nabla{\Phi_{\text{i}}}$ describes the effect of the dipole phase. For simplicity we only consider phase matching in the focal plane, so that $\boldsymbol{k_1}$ is parallel to the propagation axis and $\boldsymbol{K_\text{i}}$ is perpendicular to it. Using \eqref{eq:alphai}, assuming that the fundamental beam is Gaussian, and that the radial coordinate, $r$, is much smaller than the beam waist,
 \begin{equation}
   K_\text{i}=\frac{\partial{\Phi_{\text{i}}}}{\partial r} =-\frac{4\pi r \beta_\textrm{i}}{\lambda z_R}.
   \label{eq:K}
\end{equation}

Phase matching can therefore be achieved at certain distances from the optical axis, such that $k_q^2=q^2k_1^2+K_\text{i}^2$. This requires that the scalar quantity $\Delta k=qk_1-k_q$ is negative and, using \eqref{eq:K} together with $K_\text{i}^2\approx 2qk_1|\Delta k|$, that
\begin{equation}
r = \frac{z_\mathrm{R}}{\beta_\textrm{i}}\sqrt{\frac{q\lambda|\Delta k|}{2\pi}}.
\end{equation}
This off-axis phase matching, which depends on the electron trajectory, is characterized by a ring-like emission at the exit of the medium \cite{SalieresPRL1995}. Phase matching off-axis is easier to achieve, since $\Delta k \leq 0$. It allows for a higher intensity and ionization degree in the medium than on-axis phase matching.

\subsection{Reabsorption and other macroscopic effects}

\begin{figure}[ht]
\begin{center}
\includegraphics[]{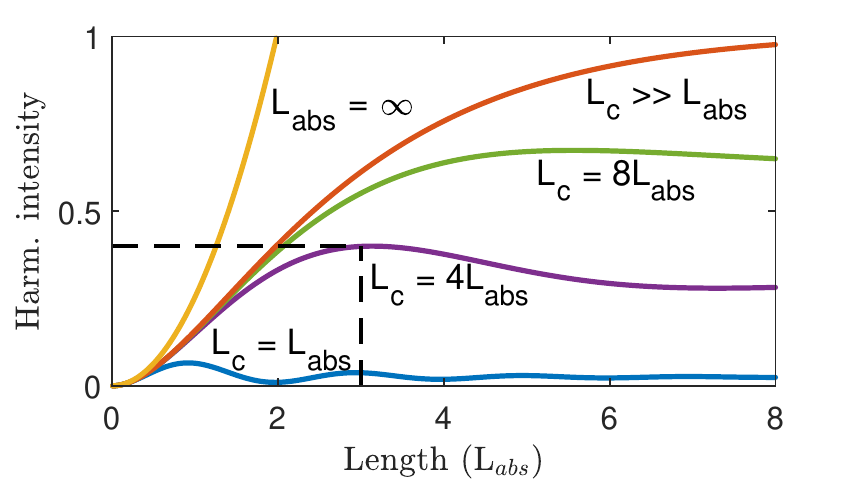}
\caption{Harmonic intensity as a function of medium length in units of $L_{\text{abs}}$ for different coherence lengths [$L_\textrm{c} = L_{\text{abs}}$, $4L_{\text{abs}}$, $8L_{\text{abs}}$, and $L_\textrm{c} \gg L_{\text{abs}}$ (blue, violet, green, red)]. The yellow line indicates the harmonic intensity calculated without absorption. The dashed black lines indicate at what medium length the maximum harmonic intensity is obtained for the case $L_\textrm{c} = 4L_{\text{abs}}$, which is related to the fit parameter $\varsigma=3$ introduced in \eqref{eq:Hyperbola}. [adapted from E. Constant (1999)]}
\label{fig:absorption2}
\end{center}
\end{figure}

Absorption of the harmonic field in the medium can be included by modifying Eq.~(\ref{eq:noabs}) as \cite{ConstantPRL1999,RuchonNJP2008}
\begin{equation}
|E_q|^2 \propto \me^{-\frac{L}{2L_\textrm{abs}}} \; \frac{\cosh\left(\frac{L}{2L_\textrm{abs}}\right) - \cos(\Delta kL)}{\Delta k^2+\frac{1}{4L_\textrm{abs}^2}}, 
\label{eq:EQsquare}
\end{equation}
where $L_\textrm{abs}$ denotes the absorption length (the inverse of the absorption coefficient at frequency $q\omega$, which is equal to $\rho \sigma_\textrm{abs}$, see Section \ref{sec:interaction}\ref{ssec:absorption}). Fig.~\ref{fig:absorption2} shows how the harmonic yield varies as a function of the medium length (in units of $L_\mathrm{abs}$) for different coherence lengths, $L_\textrm{c} = \pi/\Delta k$. The yellow line is obtained with $L_\mathrm{abs}=\infty$ and $L_\textrm{c}=\infty$. For a given coherence length, for example $L_\textrm{c} = 4L_\textrm{abs}$ (see violet curve), increasing the medium length beyond $3L_\textrm{abs}$ leads to destructive interference and limits the harmonic yield. Recently, this model has been extended to include the effects of linear density gradients along the propagation direction \cite{MajorJPB2021}.

Finally, when an intense laser beam propagates in a partially ionized medium, the radially dependent refractive index, due to the free electrons, leads to defocusing, shifting the focus towards negative $z$. For the gas densities considered in this work, Kerr induced focusing can be neglected. The peak intensity in the effective focus will then be lower than the peak intensity in vacuum. The reshaping of the pulse can lead to efficient generation through self-guiding of the driving field \cite{KimPRA2004b,MajorJOSAB2019}. Strong defocusing can also extend the phase matching cut-off due fast variations of the driving laser intensity along the propagation axis \cite{SunOptica2017}. To estimate the degree of defocusing, one can introduce a defocusing length, $L_D$, corresponding to a doubling of the diffraction limited beam divergence, defined as \cite{FillJOSAB1994,LeemansPRA1992,RaeOC1993}
\begin{equation}
 L_D = \frac{\pi c\epsilon_0m_e\omega k_\textrm{B}T}{p\eta_\textrm{fe}e^2} = \frac{\pi c\epsilon_0m_e\omega \sigma_\mathrm{abs} }{\eta_\textrm{fe}e^2}L_\mathrm{abs}. 
\end{equation}
When $\eta_\textrm{fe}=\eta_\textrm{fe}^\textrm{mac}$, $L_D= 17L_\textrm{abs}$ for the 23$^{\textrm{rd}}$ harmonic in Ar and 26$L_\textrm{abs}$ for the 69$^{\textrm{th}}$ harmonic in Ne. When $\eta_\textrm{fe}<\eta_\textrm{fe}^\textrm{mac}$, $L_D$ increases and defocusing plays a negligible role in phase matching of HHG.

\section{Simulations of high-order harmonic generation}
\label{sec:simulations}
\subsection{Method}
\label{sec:method}
\begin{figure}[ht]
\begin{center}
\includegraphics[scale=0.9]{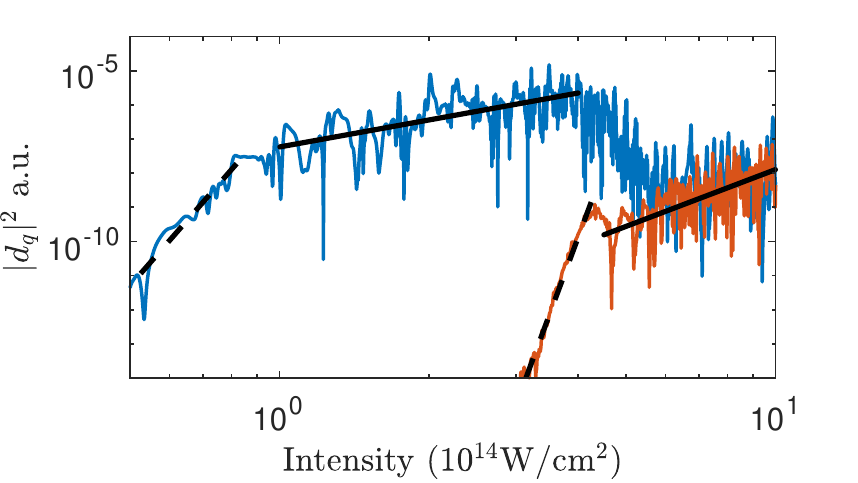}
\caption{Variation of the 23$^{\textrm{rd}}$ harmonic yield in Ar (blue) and 69$^{\text{th}}$ harmonic yield in Ne (red) with intensity, obtained by solving the TDSE for a driving wavelength of 810\,nm. The black lines represent a linear approximation $I^n$ with $n = 16.5$ (Ar) and 40 (Ne) in the cut-off region (dashed) and $n = 2.6$ (Ar) and 5.5 (Ne) in the plateau region (solid).}
\label{fig:single-atom}
\end{center}
\end{figure}

We performed numerical calculations based on the method presented in \cite{LHuillier1992}. It uses tabulated single-atom data obtained by solving the time-dependent three-dimensional Schr\"odinger equation in Ar and Ne, for a large number of peak intensity values (about five thousand) \cite{Schafer2009a}. Fig.~\ref{fig:single-atom} shows the variation of the 23$^{\textrm{rd}}$ harmonic yield in Ar (blue) and the 69$^{\textrm{th}}$ harmonic yield in Ne (red). Both curves present rapid variations as a function of intensity, resulting from interference between different electron trajectories \cite{LewensteinPRA1995}. The variation of the harmonic yields with intensity can be approximated by an $I^n$ power law with different exponents for the cut-off region and plateau \cite{KulanderJOSAB1990,LHuillierPRA1992}. The harmonic yield varies rapidly in the cut-off region, $n \approx 16.5$ for Ar and $n \approx 40$ for Ne (dashed black lines), and much slower in the plateau, $n \approx 2.6$ for Ar and $n \approx 5.5$ for Ne (solid black lines).

\begin{figure*}[ht]
\begin{center}
\includegraphics{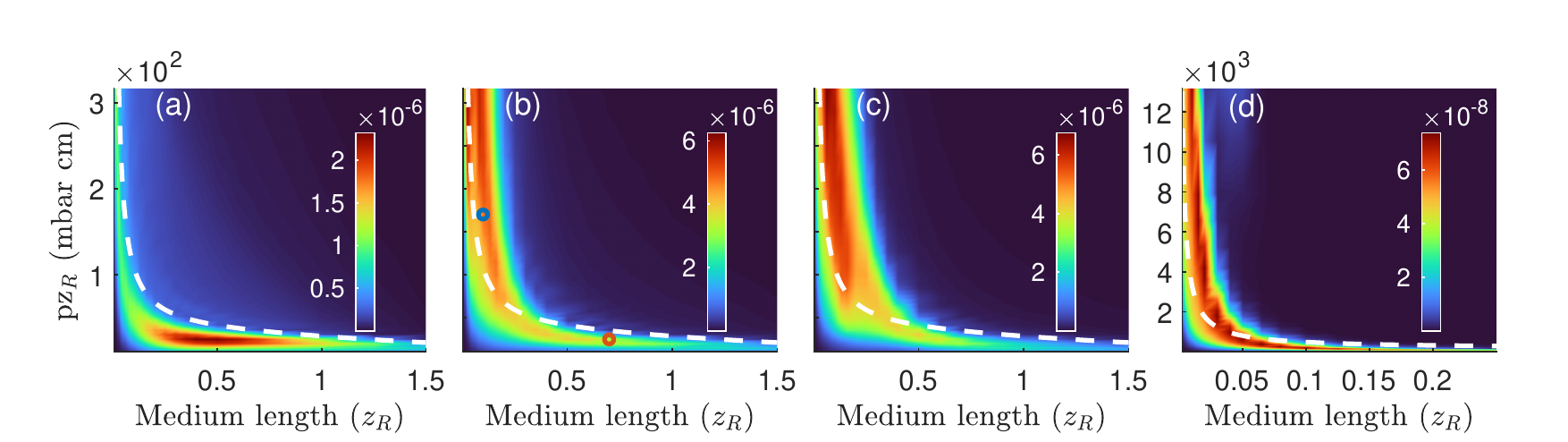}
\caption{Conversion efficiency (color scale) as a function of pressure and medium length: Ar 23$^{\text{rd}}$ harmonic (a) $I=$ 1.5 $\times 10^{14}\,$\si{\watt\per\centi\meter\squared}, (b) 2.5 $\times 10^{14}\,$\si{\watt\per\centi\meter\squared} and (c) 4.5 $\times 10^{14}\,$\si{\watt\per\centi\meter\squared}, Ne 69$^{\text{th}}$ harmonic (d) $I =$5 $\times 10^{14}\,$\si{\watt\per\centi\meter\squared}. The medium is centered at the laser focus. The dashed lines represent the prediction from \eqref{eq:Hyperbola}. The blue and red circles in (b) refer to the two cases studied further in Fig.~\ref{fig:divergenceTimeDependence}.}
\label{fig:CEq23First}
\end{center}
\end{figure*}

These data serve as input in a propagation code based on the paraxial and slowly-varying envelope approximations ({\it i.e.} it is accurate for pulse lengths longer than $\approx$ 3 cycles). For methods going beyond the slowly-varying envelope approximation see {\it e.g.} \cite{GaardeJPB2008,BrabecRMP2000,AltucciPRA2007}. We use a fundamental wavelength of 810\,nm, a pulse duration of 22\,fs, peak intensities ranging from 1.5--$4.5\times 10^{14}\,$\si{\watt\per\centi\meter\squared} in Ar and $5\times 10^{14}\,$\si{\watt\per\centi\meter\squared} in Ne. We vary the medium length and pressure. Our propagation code includes effects of reshaping of the fundamental field due to propagation in a partially ionized medium. 

There are two reasons why we use peak intensities that are larger than $I_\textrm{mac}$, thus outside the phase matching intensity window introduced in Table~\ref{tab:IntensityWindow} and illustrated in Fig.~\ref{fig:ImacTau}(a). First, the rapid variation of the single atom response with intensity strongly influences the CE, and can to some degree compensate for non-optimal phase matching. Second, HHG is a time-dependent macroscopic phenomenon, which takes place in a certain time interval and volume. The definition of $I_\textrm{mac}$ is valid only at the focus and $t=0$. While $I_\textrm{mic}$ can be considered as an intensity threshold for HHG, $I_\textrm{mac}$ is not a strict maximum intensity, but represents the highest {\it local} intensity (in time and space) at which efficient phase matching is possible.

\subsection{Simulation results}

Fig.~\ref{fig:CEq23First} shows the conversion efficiencies for the $23^\textrm{rd}$ harmonic in Ar (for several peak intensities) and for the $69^\textrm{th}$ harmonic in Ne. In both cases the medium is centered at the laser focus. The CE is represented in color as a function of medium length, in units of $z_R$, and $pz_R$, in units of \si{\milli\bar\centi\meter} (or equivalently \si{\pascal\meter}). The CE variation has a hyperbolic shape, with a maximum indicated by the dark red color, below $10^{-5}$ in Ar and $10^{-7}$ in Ne. The position of the hyperbola does not move significantly with the peak intensity, as shown for Ar in Fig.~\ref{fig:CEq23First}(a)-(c). The shape is to a high degree independent of the harmonic order and atom, as shown by comparing the results for the $23^\textrm{rd}$ harmonic in Ar and $69^\textrm{th}$ harmonic in Ne. As the intensity increases from 1.5$\times 10^{14}\,$\si{\watt\per\centi\meter\squared} to $2.5\times 10^{14}\,$\si{\watt\per\centi\meter\squared}, the maximum CE increases from $2.5\times 10^{-6}$ to 6$\times 10^{-6}$. Increasing the intensity to $4.5\times 10^{14}\,$\si{\watt\per\centi\meter\squared}, however, does not lead to significantly higher conversion efficiencies, and for even higher intensities the CE begins to decrease. Increasing the pressure further outside the range shown in Fig.~\ref{fig:CEq23First}, the maximum CE remains approximately constant, with a peak below $10^{-5}$.

This hyperbolic shape reflects the interplay between phase matching and absorption. At any pressure $p$, phase matching determines an optimum ionization degree to meet the phase matching condition $p=p_\mathrm{match}$ in \eqref{eq:pmatch5}. A sufficient peak intensity ensures that, for a wide range of pressures, this condition is reached at some point in the temporal pulse. As the pressure increases, optimum phase-matching is reached at a higher ionization degree [as illustrated in Fig.~\ref{fig:pmatch}(a)]. At the same time, absorption determines an optimum pressure-length product as shown in Fig.~\ref{fig:absorption2}. If a similar coherence length is reached, a similar optimum pressure-length product is obtained, so that the maximum CE is found on the same hyperbola. A maximum CE following such a hyperbolic relationship between pressure and length has been observed experimentally for a gas jet in \cite{CombyOE2019}.

In Fig.~\ref{fig:CEq23First}, two regimes leading to similar conversion efficiencies can be identified
\begin{itemize}
    \item[(i)]
A regime where phase matching is achieved at relatively low pressure, which requires a low ionization degree, {\it i.e.} reached early in the pulse. In this case, the neutral atom dispersion approximately cancels the sum of the Gouy phase and dipole phase contributions, {\it i.e.} $\Delta k_\textrm{at} + \Delta k_\textrm{foc} + \Delta k_\textrm{i} \approx 0$. Phase matching at a low ionization degree requires a medium with low pressure [see Fig.~\ref{fig:pmatch}(a)]. As a consequence of the low pressure, a long medium length is needed to achieve a high efficiency. In this regime, the phase matching pressure approaches its minimal value $p_0$, with
\begin{equation}
    p_0z_R = \frac{2m_\textrm{e}\omega\epsilon_0c k_{\textrm{B}}Tf_\textrm{i}}{e^2\eta_\textrm{fe}^\textrm{mac}},
    \label{eq:p0}
\end{equation}
giving $p_0z_R = 28$\,\si{\milli\bar\centi\meter} for the 23$^{\text{rd}}$ harmonic in Ar and $p_0z_R = 170$\,\si{\milli\bar\centi\meter} for the 69$^{\text{th}}$ harmonic in Ne. For the sake of generality we here include explicitly $\Delta k_\textrm{i}$ through $f_\textrm{i}$, though in Fig.~\ref{fig:CEq23First} the medium is centered at $z=0$ and $\Delta k_\textrm{i}=0$ ($f_\textrm{i} = 1$).
The CE does not vary much with the medium length, which can be explained by the fact that the process is limited by absorption (see Fig.~\ref{fig:absorption2}).
The maximum medium length is limited to values not exceeding the Rayleigh length since the dipole response far outside $z_R$ is small at the intensities considered here. This in combination with absorption leads to a lower HHG signal for lengths larger than $z_R$.
\\
\item[(ii)] A regime where HHG takes place at relatively high pressures, requiring ionization degrees close to $\eta_\textrm{fe}^\textrm{mac}$, and across a much shorter and well defined interaction length. In this case, the neutral atom dispersion approximately compensates the free electron contribution, {\it i.e.} $\Delta k_\textrm{at} + \Delta k_\textrm{fe} \approx 0$. Thus much higher pressures are required to achieve $\Delta k_\textrm{at} + \Delta k_\textrm{fe} + \Delta k_\textrm{foc} = 0$. In this regime the efficiency depends strongly on the medium length but not much on the pressure, provided the pressure is sufficiently high ($p\gg p_0$). 
\end{itemize}

\subsection{Analytic form of the hyperbola}

To parametrize the hyperbolic shape exhibited by the simulations, we assume that HHG is absorption-limited \cite{ConstantPRL1999} with $L=\varsigma L_\textrm{abs}$, where $\varsigma=3$ is a fit parameter depending on the achieved coherence length (see dashed black line in Fig.~\ref{fig:absorption2}):

\begin{equation}
    pz_R\frac{L}{z_R} = \frac{\varsigma k_BTf_\textrm{i}}{\sigma_\textrm{abs}}
    \label{eq:PressureLength}
\end{equation}

This suggests a (partial) explanation in which the medium should not be too short (which will make full coherent build-up impossible) but also not too long (which will not lead to any additional coherent build-up, but leads to deteriorating effects such as decoherence or plasma effects).

Including the phase matching pressure for the case of $\eta_\textrm{fe}=0$, introduced in \eqref{eq:p0} we propose the following equation for the hyperbola
    \begin{equation}
    (p-p_0)L=\frac{\varsigma k_\textrm{B}Tf_\textrm{i}}{\sigma_\textrm{abs}}.
    \label{eq:Hyperbola}
    \end{equation}
    The factor $f_\textrm{i}$ can be chosen to be constant, equal to 1, for a medium centered at the focus. Alternatively, we can use $f_\textrm{i}=f_\textrm{i}(L/2)$, since the XUV radiation emitted at the end of the medium ($z=L/2$) has experienced the least reabsorption. This leads to a small correction to the hyperbolic equation, with a slightly tilted ``horizontal'' branch. 
    \eqref{eq:Hyperbola}, represented by the dashed white line in Fig.~\ref{fig:CEq23First} for $\varsigma = 3$, agrees with the simulations in both Ar and Ne for harmonics in the plateau region. The hyperbolic behavior, which is the main finding of this work, is remarkably universal, {\it i.e.} independent of the peak intensity, harmonic order, generating gas, and focusing geometry through the scaling laws presented in Section \ref{sec:principles}\ref{sec:scaling}. 

\subsection{Spatial and temporal profile of the emitted XUV radiation}
\begin{figure}[ht]
\begin{center}
\includegraphics[scale=0.9]{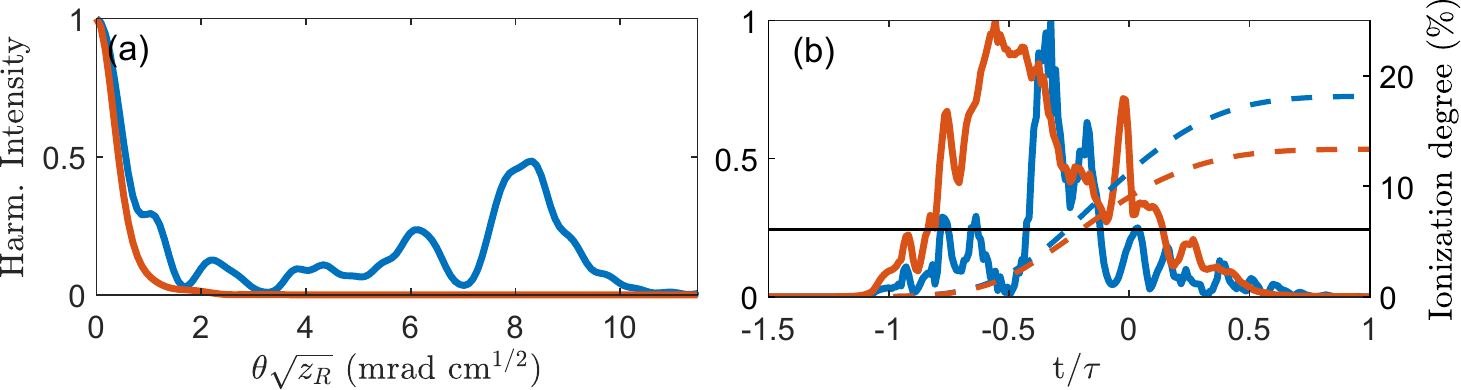}
\caption{(a) Normalized 23$^\textrm{rd}$ harmonic intensity in Ar as a function of divergence angle in the far field and (b) as a function of time, with $t=0$ at the peak of the fundamental pulse, for the cases of high pressure (blue) and low pressure (red). The average ionization degree on axis in the medium as a function of time is shown for high pressure (dashed blue) and low pressure (dashed red). The horizontal black line indicates the critical ionization degree $\eta_\textrm{fe}^{\textrm{mac}}.$}
\label{fig:divergenceTimeDependence}
\end{center}
\end{figure}

The spatial profile of the emitted XUV radiation is shown in Fig.~\ref{fig:divergenceTimeDependence}(a) using $I=2.5\times10^{14}\,\si{\watt\per\centi\meter\squared}$ for the 23$^{\text{rd}}$ harmonic in Ar for $(p,L)=(25$\,\si{\milli\bar\centi\meter}, 0.7\,$z_R$) and $(p,L)=(170$\,\si{\milli\bar\centi\meter}, 0.1\,$z_R$) corresponding to regime (i) and (ii), respectively (indicated in Fig.~\ref{fig:CEq23First}(b) as the red and blue circles). As can be seen, the harmonic is much less divergent in regime (i) being confined to a solid angle of $\theta \sqrt{z_R} < $ 2\,\si{\milli\radian\centi\meter}$^{1/2}$. In contrast, the spatial profile in regime (ii) is more irregular with a strong off-axis contribution, possibly due to off-axis phase matching, as discussed in Section \ref{sec:principles}\ref{sec:offAxis} and/or due to the contribution of the long trajectory.

Fig.~\ref{fig:divergenceTimeDependence}(b) shows the temporal profile of the XUV radiation exiting the medium, together with the average ionization degree on axis. In both regimes the generation takes place mostly prior the peak of the IR pulse. The emission is, however, much more transient in regime (ii) than in regime (i), taking place at a later time when the ionization degree is close to the critical ionization degree, indicated by the horizontal black line [see also Fig.~\ref{fig:pmatch}(a)] \cite{HeylJPBAMOP2017}. The higher ionization degree in regime (ii) is a consequence of a higher average intensity in the shorter medium.

\begin{figure}[ht]
\begin{center}
\includegraphics[]{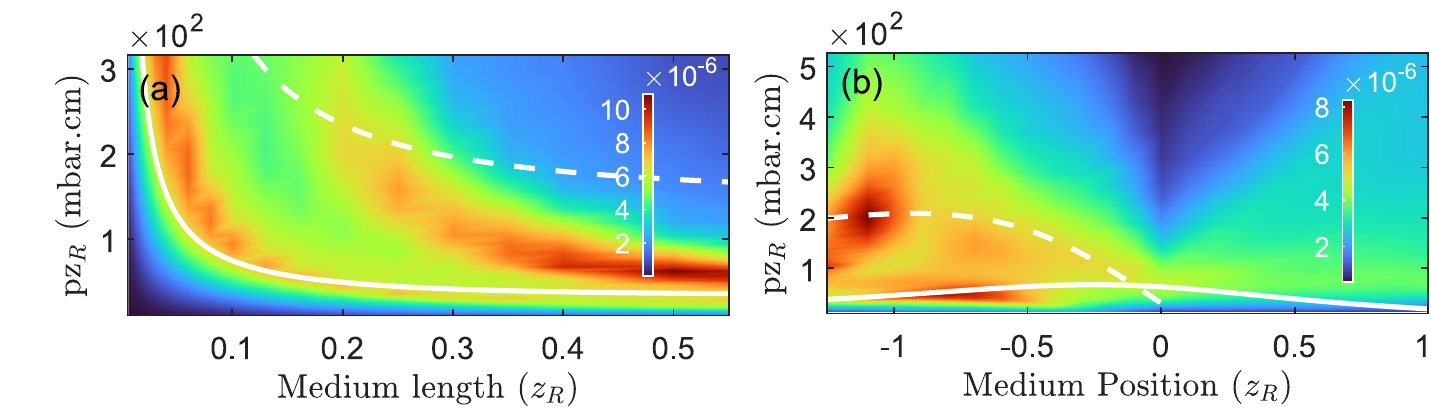}
\caption{ Conversion efficiency (color scale) for the 23$^{\text{rd}}$ harmonic in Ar as a function of (a) $pz_R$ and medium length, with the medium centered at z = $-z_R$ and (b) $pz_R$ and medium position, for a medium of length $L=0.2z_R$. The solid (dashed) white lines indicate how \eqref{eq:Hyperbola} varies for the short (long) trajectory. The intensity at the center of the medium is 2.5$\times 10^{14}\,$\si{\watt\per\centi\meter\squared}.}
\label{fig:minus1zr}
\end{center}
\end{figure}
\subsection{Variation of the medium position}
Finally, we vary the position of the medium relative to the laser focus, which is known to affect the conversion efficiency, \cite{SalieresPRL1995,MajorO2021} and the position of the harmonic focus \cite{WikmarkPNAS2019,QuintardSA2019,HoflundUS2021}. In Fig.~\ref{fig:minus1zr}(a), we show the CE of the $23^\textrm{rd}$ harmonic in Ar (color scale) as a function of medium length and $pz_R$ for a medium centered at $z = -z_R$, {\it i.e.} before the laser focus. The laser intensity at the center of the medium is 2.5$\times 10^{14}\,$\si{\watt\per\centi\meter\squared}.
We observe two hyperbolic shapes with similar CE, converging to different pressures, approximately 26\,\si{\milli\bar\centi\meter} and 58\,\si{\milli\bar\centi\meter}. The hyperbolic equation depends on the different trajectories through $f_i$ [see \eqref{eq:pmatch4} and Fig.~\ref{fig:pmatch}(b)]. The equation for the short trajectory (solid white line) follows the simulation results very well, while that of the long trajectory (dashed line) follows the trend of the upper feature at high pressure. 

Fig.~\ref{fig:minus1zr}(b) shows how the CE varies as a function of medium position, $z$ and $pz_R$ for a constant medium length of $L = 0.2z_R$. The solid (dashed) white line shows how $pz_R$, extracted from \eqref{eq:Hyperbola}, varies with $z$ for the short (long) trajectory. At $z = 0$ the long and short trajectory phase match at the same pressure, since there is no dipole phase contribution, leading to a region of high CE which is confined to a narrow range of pressures. For $z_ > 0$, (medium after the laser focus) $f_\ell$ becomes negative, meaning that phase matching can only be achieved for the short trajectory. For $z < 0$ the CE increases, reaching a maximum at $z = -0.75z_R$ (short trajectory) and $z = -1.1z_R$ (long trajectory). These maxima correspond to optimized generating positions, such that the laser- and dipole-phase-induced wavefronts compensate each other \cite{WikmarkPNAS2019,QuintardSA2019}, which are different for the two trajectories. In the interval $-1.25z_R < z < -0.5z_R$ the CE displays a double peak structure, with the largest splitting at $z = -z_R$. The regions of high CE agree quite well with the phase matching pressure predictions for the long (dashed line) and short (solid line) trajectories. This analysis suggests that the upper hyperbolic feature in Fig.\ref{fig:minus1zr}(a) corresponds to phase matching of the long trajectory, while the lower one to that of the short trajectory. 

\section{Examples}
\label{sec:examples}
The initial motivation for this work was to find a ``recipe'' for phase-matched, efficient HHG, {\it i.e.} to suggest a best choice of geometry, medium length and gas pressure for a given set of laser parameters. We find that there is no unique best solution, but several, with advantages and disadvantages. We illustrate this below with two examples.

Given a driving pulse energy $E=1\,$mJ, duration $\tau=22\,$fs FWHM and wavelength of $810\,$nm, we want to optimize the number of photons of the 23$^\textrm{rd}$ harmonic in Ar, with (i) and without (ii) the additional requirement of a high spatial quality.

\subsection{High conversion efficiency and spatial quality} 
A high spatial quality requires to be in regime (i), {\it i.e.} on the horizontal branch of the hyperbola [see Fig.~\ref{fig:divergenceTimeDependence}(a)]. We choose a focal intensity equal to 2.5$\times 10^{14}\,$\si{\watt\per\centi\meter\squared}, such that $I \ge I_\textrm{mic}$ over a large volume in the medium. Above this intensity the ionization increases and the CE in regime (i) saturates. Assuming Gaussian optics, the Rayleigh length must be chosen to be
\begin{equation}
        z_R = \frac{4\sqrt{\ln{2}}E}{\sqrt{\pi}\tau \lambda I} = 0.042\,\si{m},
\end{equation}
which requires a focal length $f=\sqrt{\pi z_R/4\lambda}D= 200D$, where $D$ is the unfocused beam diameter. The pressure should be approximately 7\,mbar according to \eqref{eq:p0} and \eqref{eq:Hyperbola}. The medium length is not a sensitive parameter in this regime. However, the simulations [see Fig.~\ref{fig:CEq23First}(a-c)] indicate that a length equal to $0.5z_R\approx 2.1\,$cm is appropriate. The obtained CE is $4\times 10^{-6}$. The beam is well collimated, with a smooth spatial profile and a divergence less than 1\,mrad.

\subsection{High conversion efficiency} 
We now target regime (ii), {\it i.e.} the vertical branch of the hyperbola. We choose a higher focal intensity, equal to 4$\times 10^{14}\,$\si{\watt\per\centi\meter\squared}, to reach an intensity $I \approx I_\textrm{mac}$ over a large volume in the medium. This implies a Rayleigh length equal to $z_R = 0.026\,\si{m}$, and $f= 160 D$. The pressure-length product is equal to 8\,\si{\milli\bar\centi\meter} [see \eqref{eq:PressureLength}]. As shown in the simulations, the CE increases only slightly when increasing the pressure while decreasing the medium length. In many experimental situations, either the pressure or the length will be determined by technical limitations ({\it e.g.} highest backing pressure or shortest possible medium length in the case of a gas jet). Using a pressure equal to 220\,mbar the length of the medium should be of the order of 0.35\,mm. The obtained CE is approximately 8$\times10^{-6}$. The harmonic beam has a poor spatial quality with many annular spatial structures, and a divergence of the order of 6\,mrad [see Fig.~\ref{fig:divergenceTimeDependence}(a)].

\section{Conclusion}
\label{sec:discussion}
In summary, we have reviewed the main physical principles of high-order harmonic generation in gases, with emphasis on the macroscopic aspects. Using numerical simulations we have calculated the CE as a function of various parameters (pressure, intensity, medium length, position of medium relative to the laser focus, focusing geometry and trajectory) in Ar and Ne.

 Efficient generation follows approximately a hyperbolic shape [$(p-p_0)L\propto 1/\sigma_\textrm{abs}$, see \eqref{eq:Hyperbola}]. The pressure must be larger than $p_0$ (see \eqref{eq:p0}), while $L$ should be less than approximately a Rayleigh length. While the conversion efficiency remains approximately constant along this hyperbola, the properties of the harmonics, however, strongly depend on the choice of $(p,L)$. At low pressure, the maximum CE does not depend much on the medium length and the generated XUV radiation is well collimated. In contrast, at high pressure, the maximum CE is relatively pressure-independent, and the generated radiation is much more divergent and emitted more transiently, later in the pulse. Therefore, the best choice of $(p,L)$ depends on whether a high spatial quality, high CE or a flexible experimental setup is prioritized for a specific application. Finally, moving the medium approximately one Rayleigh length before the laser focus leads to an enhanced CE. In this case, optimal phase matching of the long and short trajectories occurs at different pressures and lengths, leading to two hyperbolas.

In conclusion, for any given focusing geometry, harmonic order and generating gas, efficient generation is possible for a variety of gas pressures and medium lengths.
This explains the large versatility in gas target design for efficient HHG and will provide design guidance for future high-flux XUV sources.

\section*{Disclosures}
\noindent\textbf{Disclosures.} The authors declare no conflicts of interest.

\section*{Acknowledgements}
The authors would like to thank Kenneth J. Schafer and Mette B. Gaarde for help regarding the numerical simulations. The authors acknowledge support from the Swedish Research Council (2013-8185, 2016-04907), the European Research Council (advanced grant QPAP, 884900) and the Knut and Alice Wallenberg Foundation. AL is partly supported  by the Wallenberg Center for Quantum Technology
(WACQT) funded by the Knut and Alice Wallenberg foundation. L.R acknowledges support from Ministerio de Educación, Cultura y Deporte (FPU16/02591).

\bibliography{Ref_lib}

\end{document}